# High strength self-healable supercapacitor based on supramolecular polymer hydrogel with upper critical solubility temperature.

Roman Elashnikov[a]*, Olena Khrystonko[a], Tereza Jilková[a], Silvie Rimpelová[b], Zdenka Kolská[c], Václav Švorčík[a], Oleksiy Lyutakov[a]

[a]Department of Solid State Engineering, University of Chemistry and Technology Prague, Technická 5, 166 28, Prague 6, The Czech Republic,

[b]Department of Biochemistry and Microbiology, University of Chemistry and Technology Prague, Technická 5, 166 28, Prague 6, The Czech Republic

[c]Faculty of Science, J. E. Purkyně University in Ústí nad Labem, České Mládeže 8, 400 96 Ústí nad Labem, Czech Republic

*corresponding author: roman.elashnikov@vscht.cz

## Abstract

Here, we report poly(N-acryloylglycinamide-*co*-vinyltriazole) p(NAGA-*co*-VTZ) supramolecular polymer hydrogel doped with activated polypyrrole nanotubes (acPPyNTs) as a high-strength self-healable material for supercapacitors. First, the p(NAGA-*co*-VTZ) hydrogel films were synthesized by photopolymerization of N-acryloylglycinamide and 1-vinyl-1,2,4-triazole without any cross-linkers. Scanning electron microscopy and mechanical tests showed that initial monomer concentration strongly affects both hydrogel microstructure and resulted mechanical properties. The hydrogels demonstrated self-healing ability through hydrogen bonding at the temperatures above upper critical solubility temperature, excellent mechanical properties (0.9 MPa), large stretchability (1300 %) and cut resistance. Next, as active material for electrochemical double layer capacitors (EDLC) carbonized and ethanol/KOH activated polypyrrole nanotubes (acPPyNTs) were prepared. Symmetric self-healable supercapacitor employing p(NAGA-*co*-VTZ) hydrogel, acPPyNTs and aqueous 3M KCl solution was assembled. Cyclic voltammetry, galvanostatic charge-discharge measurements showed that the prepared device gave a specific capacitance of 282.62 F $g^{-1}$ at 0.2 A $g^{-1}$ and high areal capacitance of 316.86 mF $cm^{-2}$ at scan rate of 10 mV $s^{-1}$. The supercapacitor operates over wide voltage window (0-1.2 V) and provides excellent cyclic performance with capacitance retention of 97 % after 10 000 cycles and 94 % after self-healing. Overall, the prepared self-healable supercapacitor appears to have considerable potential as high-performance energy storage device.

# 1. Introduction

Wearable electronics have received much attention because of their ability to provide useful insights into the performance and health of people.[1] The development of many new soft electronic devices for on-body-sensing, recording electrical and physiological signals has been reported.[2,3] In particular, such devices can provide non-invasive health monitoring, detect metabolites, nutrients or even indicate warfare agents.[4,5] Real-time monitoring can collect and provide information about health, improve the management of chronic diseases, and alert the user or medical professionals of abnormal or unforeseen situations. Furthermore, recent research showed that such devices can be used for energy generation, storage and utilization.[6,7] In particular, energy can be stored using flexible supercapacitors which can provide high power output in short time.[8–11] However, even minor mechanical damages to such sift and flexible devices lead to a significant deterioration or irreversible loss of their properties and functionality.[12,13] To address this, significant attention have been devoted to increasing of the mechanical properties and performance of flexible supercapacitors and introduction of self-healing functionality.[14]

As self-healing materials, polymers and polymer-based composites gain significant attention because of their light weight, flexibility and stretchability.[15] One of the most popular options for self-healable devices represents polyvinylalcohol (PVA), due to its chemical stability and electrochemical inertness.[16] The performance of PVA-based supercapacitors is strongly dependent on additives, such as 2D materials, conducting polymers or carbon nanotubes, using for creation of hybrid materials and the fabrication of supercapacitors.[17] The main drawback of such approach is low mechanical properties of PVA, dehydration, worse self-healing performance in the presence of inorganic ions, which can disrupt hydrogen bonds.[16,18] To another class of hydrogen-bonding alginate-based supercapacitors were reported.[19] Again, the mechanical properties of such materials leave much to be desired and the self-healing performance is highly dependent on the leaching of cross-linking ions. Another alternative is based on acrylic monomers, connected by multifunctional cross-linker *N,N*-bis(acryloyl)cystamine and doped with metal nanoparticles or conducting polymers.[20] These materials demonstrated relatively good mechanical properties, but are limited by the necessity of using metallic nanoparticles for cross-linking. To the best of our knowledge, in the field of self-healable supercapacitors less attention has been given to polymers with upper-critical solubility temperature. (UCST). Such materials provide dynamic hydrogen bonding, can have outstanding for hydrogels mechanical properties do not require additional cross-linking agents can be repaired by heating and have good processability potential.

This work demonstrates potential of acPPyNTs doped poly(N-acryloylglycinamide-*co*-vinyltriazole) (p(NAGA-*co*-VTZ)) hydrogel as self-healable material for supercapacitor devices. This material has wide potential window, high capacitance, energy density, cyclic performance, good mechanical properties, and temperature-induced self-healing functionality.

## 2. Results and discussion

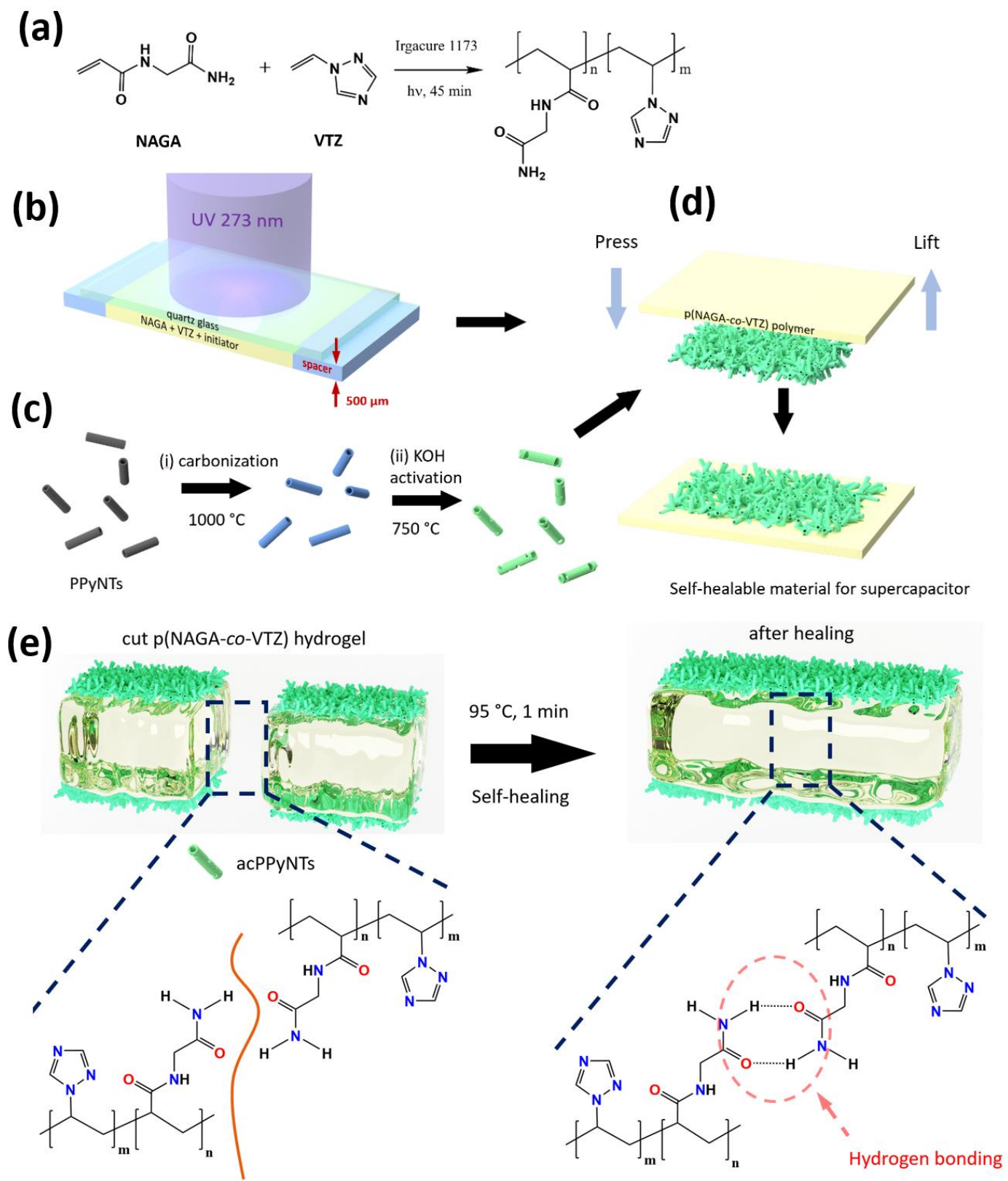

**Scheme 1** preparation steps of self-healable supercapacitor based on p(NAGA-*co*-VTZ) hydrogel with upper critical solution temperature: copolymerization of N-acryloyl glycinamide and 1-vinyl-1,2,4-triazole (a), synthesis setup for copolymerization (b), preparation, carbonization and activation of polypyrrole nanotubes (PPyNTs) (c), deposition of activated PPyNTs on p(NAGA-*co*-VTZ) hydrogel film (d), heating-induced self-healing mechanism of p(NAGA-*co*-VTZ) hydrogel (e).

**Scheme 1a-d** represents fabrication process of self-healable p(NAGA-*co*-VTZ) hydrogel supercapacitor. First, the polymer hydrogel was prepared by copolymerization of N-acryloyl glycinamide and 1-vinyl-1,2,4-triazole using photoinitiator. The gel was prepared by UV-photopolymerization using 500 μm spacers, which enable us to control the film thickness. Activated polypyrrole nanotubes (PPyNTs) were used as a model for charge storage. Polypyrrole (PPy) represents an attractive option as template for the preparation of high surface area nitrogen-doped carbon. The ability of CPs to retain their morphology after carbonization is of outmost importance, because after such treatment the nanostructures remain their shape, retaining high specific surface area.[21] Furthermore, it was reported that doping of the carbon with nitrogen can not only the conductivity of carbon-based materials but also increase charge storage and release performance.[22] In addition, the specific surface area of such carbonized structures can be significantly enhanced by steam or potassium hydroxide activation.[23] Thus, PPyNTs, prepared by oxidation polymerization, were subsequently carbonized and activated using potassium hydroxide aqueous solution with ethanol. Different concentration of KOH was used to activate cPPyNTs to evaluate the efficiency of the process. Then acPPyNTs were placed on glass and the p(NAGA-*co*-VTZ) hydrogel film was pressed and slowly peeled by hand to transfer acPPyNTs onto polymer hydrogel. Next, such films were soaked in electrolyte solution to saturate both the membrane and supercapacitor was assembled. The fabricated material possesses self-healing ability due to multiple hydrogen bonding domains. **Scheme 1e** shows self-healing mechanism of p(NAGA-*co*-VTZ) hydrogel, namely, cut of the prepared material doped with acPPyNTs and temperature-induced hydrogen bonds formation between amide groups of N-acryloyl glycinamide after mechanical damage.

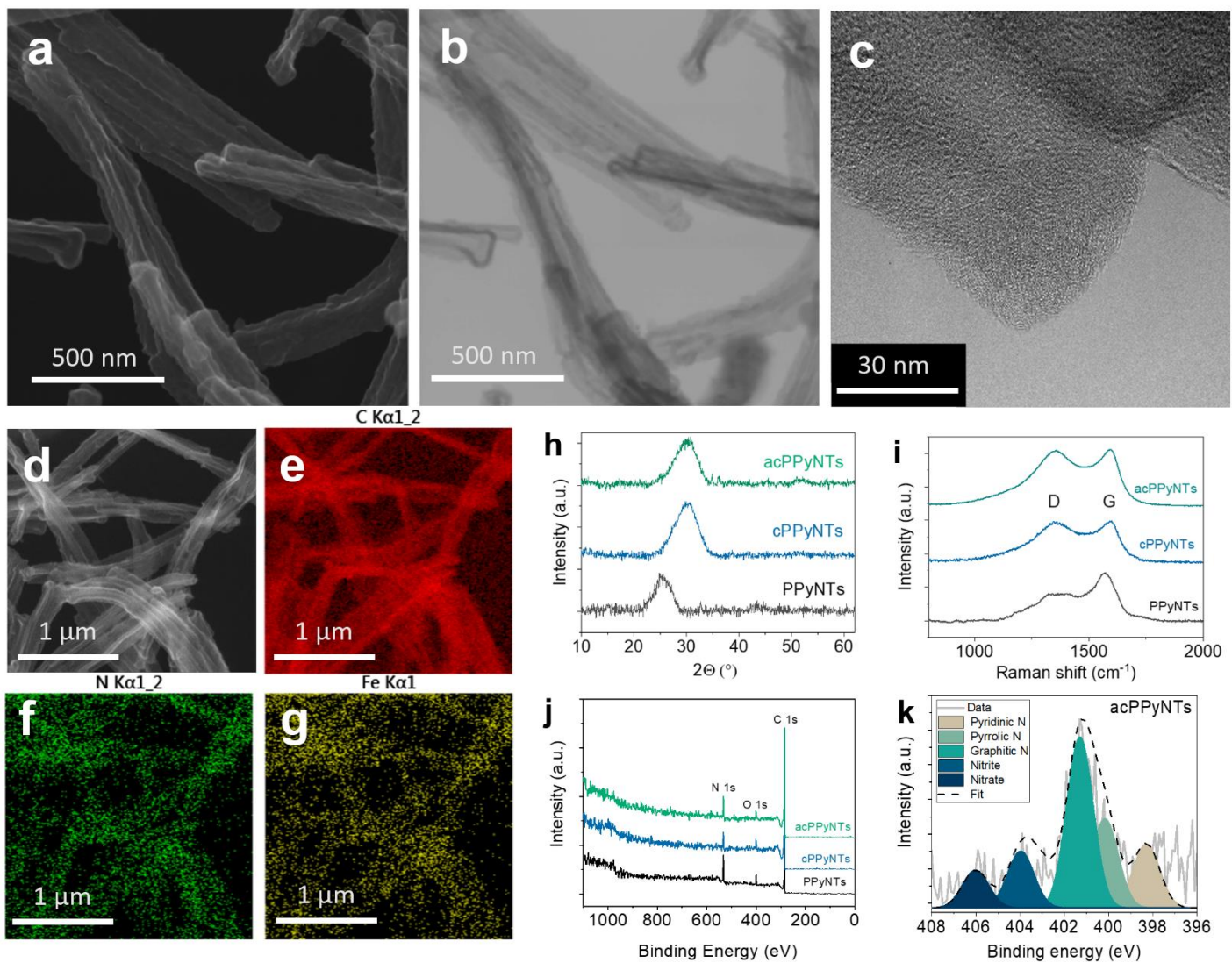

**Figure 1** SEM (a), STEM (b) and HRTEM (c) images of KOH-activated carbonized polypyrrole nanotubes (acPPyNTs). SEM (d) and SEM-EDX elemental mapping images of C (e), N (f), Fe (g). XRD (h), Raman (i) and XPS (j) spectra of polypyrrole nanotubes (PPyNTs), carbonized polypyrrole nanotubes (cPPyNTs) and acPPyNTs. high-resolution XPS spectra (k) of deconvoluted N1s peak of acPPyNTs.

First, the morphology of the prepared PPyNTs was analysed by scanning electron microscopy. SEM and STEM analysis of as-prepared PPyNTs showed narrow and uniform nanotube morphology with the average diameter of 95 nm and the length of a single nanotube was in the range from 1 to 12 μm (**Figure S1a**). The average wall thickness of the PPyNTs was 18 nm. After carbonization, the average diameter of the nanotubes slightly decreased to 90 nm and the average wall thickness was 16.4 nm (**Figure S1b**). As is evident, the PPyNTs remain their shape after carbonization and the surface morphology of the carbonized PPyNTs became smoother. Further, KOH activation of PPyNTs does not significantly change the size of the single PPyNTs, but slightly affect their size distribution, which can be related to the activation process and process operation during sample preparation (**Figures 1a, 1b**).

Preserving the morphology of PPyNTs is important because maintaining the interconnected porous structure can decrease ion transport resistance.[24] HRTEM analysis showed amorphous carbon without crystalline inclusions (**Figure 1c**). The SEM-EDX analysis performed to exclude the presence of oxides revealed a homogeneous distribution of carbon nitrogen and iron in activated PPyNTs (**Figure 1e-g**). BET analysis showed that as synthesised PPyNTs have 62.9 ± 3.1 m$^2$ g$^{-1}$ and pores volume 0.095 ± 0.004 cm$^3$ g$^{-1}$. After carbonization at 1000 °C their specific surface area decreased to 49.1 ± 2.1 m$^2$ g$^{-1}$, which corresponds to the changes of surface morphology, observed by SEM (**Figures S1A, B**). After the activation, nanotubes specific surface area (SSA) increased significantly to 411.1 ± 11.1 m$^2$ g$^{-1}$ with a pore volume 0.29 ± 0.011 cm$^3$ g$^{-1}$, indicating successful activation of PPy-based nanotubes. High surface area in combination with N and Fe doping are of potential interest for subsequent supercapacitor application.

Next, chemical and structural analysis of the prepared nanotubes was performed. X-ray diffraction (XRD) analysis, performed to exclude the presence of iron oxide, indicates that all iron is in the atomic form (**Figure 1h**). No characteristic Fe$_2$O$_3$ or Fe$_3$O$_4$ peaks were observed on XRD spectra. A typical broad diffraction peak for PPyNTs can be observed at 2θ around 25° suggesting amorphous structure of PPyNTs. For carbonized nanotubes as well as for KOH-activated nanotubes can be observed shift of the diffraction peak towards higher values of 30°. Raman spectroscopy of cPPyNTs and acPPyNTs also demonstrated typical for carbon structures peaks with 1353 and 1595 cm$^{-1}$ maxima corresponding to disordered (D) and graphitic (G) bands, respectively (**Figure 1i**). The spectra were deconvoluted using Origin2018 software and Lorentzian peaks. The degree of crystallinity was calculated as the integrated intensities of the D and G bands ($I_D/I_G$), which was 2.71 for cPPyNTs and 2.78 for acPPyNTs. In turn, XPS analysis showed the presence of carbon, nitrogen, oxygen and small amount of iron in all samples (**Figure 1j**). After carbonization and activation, the XPS analysis showed the sample composition was C (92.5 %), N (2.9 %), O (4.4 %) and Fe (0.2 %). High-resolution N1s deconvoluted peaks revealed five different nitrogen energy, namely 398.3 eV related to pyridinic (=N–H), pyrrolic 400.1 (–N–H) graphitic (=N$^+$–H) and N-oxides (nitrite and nitrate) at 403 and 406 eV nitrogen peaks. From XPS measurements it is obvious that the amount of graphitic nitrogen is prevails in acPPyNTs (Figure 1k). residual potassium was not revealed by XPS.

Separately, p(NAGA-*co*-VTZ) hydrogel preparation was optimized. Different ratio of monomers was used to achieve a stable film. It was found that p(NAGA-*co*-VTZ) hydrogel films can be prepared in the range of monomer concentration from 9 to 37 %. This observation is in good agreement with the literature, because N-acryloylglicineamide can not form gels in water solution below a certain concentration.[25] The highest concentration corresponded to better mechanical properties and film stability. Further increase of the monomer concentration led to the formation of non-homogeneous hydrogel, with a wide pore sizes distribution. The morphology of the prepared p(NAGA-*co*-VTZ) hydrogel, depending on initial monomer concentration, analysed by scanning electron microscopy (SEM), is given in SI. SEM showed that the increase in monomer concentration led to the reduction in p(NAGA-*co*-VTZ) hydrogel pore size (**Figure S2**).

In the next step, chemical composition of the polymer hydrogel and its chemical composition was analysed. Fourier transform infrared analysis (FTIR), performed to confirm the chemical composition of the prepared hydrogel, showed the presence of both monomers (**Figure 2b**). FTIR spectra copolymer p(NAGA-*co*-VTZ) showed characteristic amide peaks at 1648 cm$^{-1}$ represents vibration of C=O stretching in -CO-NH$_2$- group and peak at 1629 cm$^{-1}$ typical for C=O stretching and NH$_2$ deformation. The peak at 1550 cm$^{-1}$ is connected to N-H stretching in triazine ring and 1509 cm$^{-1}$ to C=N stretching in triazine ring. Peaks 1423 and 1400 cm$^{-1}$ are characteristic for C-H bending in -CHO group and C-N stretching in -CO-NH$_2$- group.

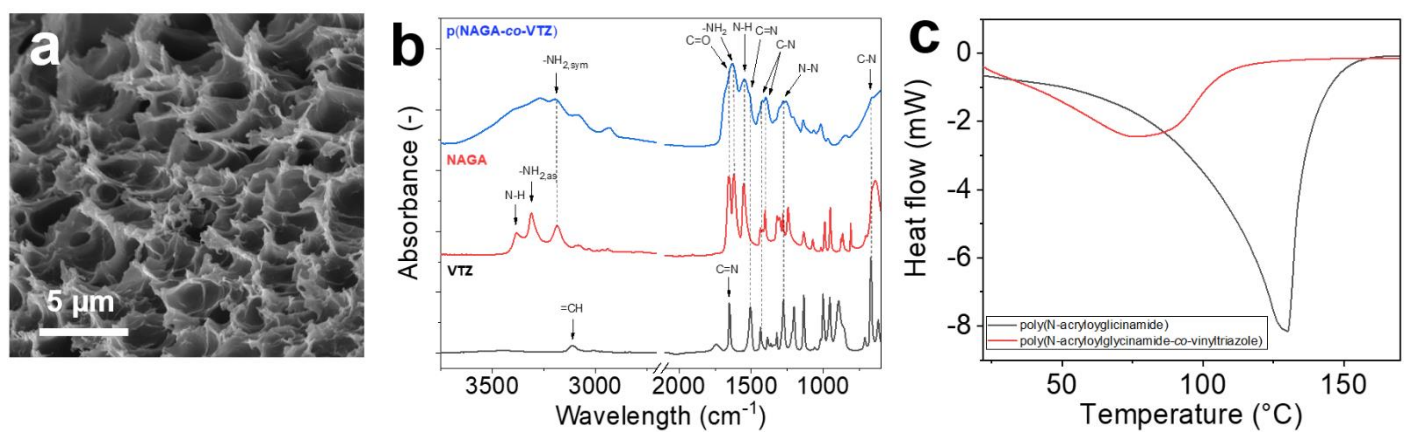

**Figure 2** SEM image of surface hydrogel morphology, used for supercapacitor preparation (a), FTIR spectra of dry p(NAGA-*co*-VTZ) polymer, N-acryloyl glycinamide and 1-vinyl-1,2,4-triazole monomers (b), differential scanning calorimetry analysis of pristine PNAGA polymer compared to p(NAGA-*co*-VTZ) copolymer (c).

Peaks at 3268 and 3195 cm$^{-1}$ related to asymmetric and symmetric stretching of primary amide. The peak at 3079 cm$^{-1}$ relates to =C–H stretching of triazin ring, which in this spectre is slightly shifted in comparison to the FTIR spectre of VTZ. The peak at 2931 cm$^{-1}$ is related with asymmetric stretching of –CH$_2$ bond in –C–CH$_2$– group. Peak at 1276 cm$^{-1}$ corresponds to N-N stretching in the triazine ring and the peak 665 cm$^{-1}$ is assigned to C-N stretching out of the triazine ring.

To analyse the UCST of the prepared p(NAGA-*co*-VTZ) hydrogel and to estimate the temperature required for self-healing, differential scanning calorimetry (DSC) was used. DSC indicated a strong shift of melting point of pristine PNAGA hydrogel from 130.1 °C to 75 °C compared to p(NAGA-*co*-VTZ) copolymer, respectively (**Figure 2c**). The melting temperature of PNAGA polymer is in good agreement with the literature.[26] Melting point temperature of PNAGA polymers can be strongly affected by impurities or synthesis route. As is evident from **Figure 2c** the copolymerization of NAGA monomer with 1-vinyl-1,2,4-triazole have a strong effect on melting point. It was observed not only change in the melting temperature, but also a broadening of the melting peak, allowing for the use of a wide temperature range for self-healing. Based on DSC analysis the temperatures above 75°C was selected for self-healing of the polymer.

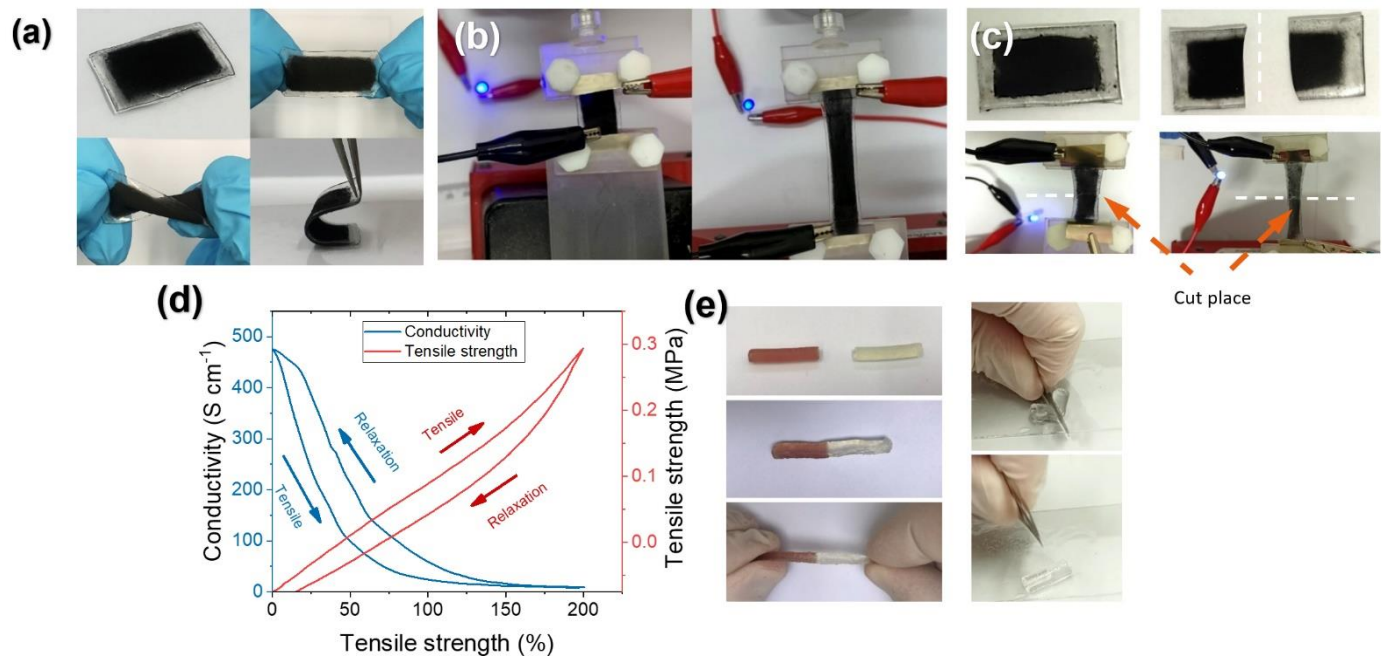

**Figure 3** The optical images showed that the p(NAGA-*co*-VTZ) hydrogel with deposited acPPyNTs could endure multiple deformations, including stretching, twisting and bending (a), set-up for measurement of cyclic tensile and conductivity tests of acPPyNTs doped p(NAGA-*co*-VTZ) hydrogel connected to blue LED, indicating changes of surface conductivity upon mechanical strength (b) cut and healing of the samples and demonstration of electromechanical properties of the hydrogel after self-healing (c), electro-mechanical tests of the p(NAGA-*co*-VTZ) hydrogel doped with acPPyNTs (d), self-healing of hydrogel pieces prepared separately and resistance of the prepared hydrogel tube to cut (e).

Optical images (**Figure 3a**) show p(NAGA-*co*-VTZ) hydrogel films covered with acPPyNT. During different deformations, including stretching, twisting and bending the layer of carbon nanotubes remain stable. Mechanical tests showed a tensile strength and stretchability depends on initial monomer concentration (**Figure S3**). The best mechanical properties were achieved in the case of the highest concentration at 37 % wt. tensile strength and elongation was observed. Excellent of 0.9 MPa and large stretchability 1300 %. In the next step, electromechanical measurements were performed in order to evaluate acPPyNYs layer continuity on the surface of p(NAGA-*co*-VTZ) hydrogel (**Figure 3b**). The photos display changes in the brightness of the LED connected to the circuit containing the acPPyNTs doped p(NAGA-*co*-VTZ) hydrogel. Even under a stretch of more than 300 %, the LED continues to emit light. To show self-healing ability, p(NAGA-*co*-VTZ) hydrogel film with deposited acPPyNTs was cut in two pieces and than healed during 1 min at 95 °C. As is evident from **Figure 3c**, the hydrogel was not only successfully healed but also remain surface conductivity and good mechanical properties. The results of simultaneous tensile stress surface conductivity of non-damaged sample are presented on **Figure 3d**. From these graphs, it is evident that the gel can be reversibly stretched while maintaining its mechanical properties. Hysteresis loops were evident in the tensile cycles, which that the hydrogel retain network structure. The hydrogel conductivity decreases from 474 S cm$^{-1}$ to 10 S cm$^{-1}$.

The prepared p(NAGA-*co*-VTZ) hydrogel showed not only good tensile strength but also cut-resistance. The evidence of cut resistance is presented on **Figure 3e**, it is shown on that the prepared p(NAGA-*co*-VTZ) hydrogel tube could not be easily cut with razor blade. The example prepared polymer hydrogel with UCST behaviour showed self-healing properties at 95 °C during 1 min under slight compression in glass tube to ensure a better contact. Two pieces of hydrogel were separately synthesized. One of the samples was prepared with a small amount of dye (red one). The hydrogels demonstrated good self-healing ability (**Figure 3f**). As can be seen from tensile test the hydrogels remain 92 % of their initial strength.

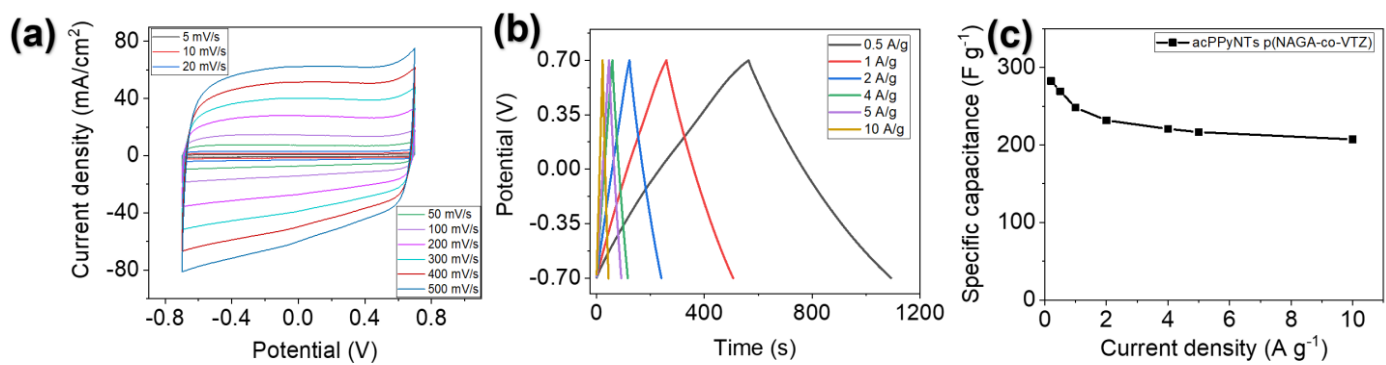

**Figure 4.** CV curves of acPPyNTs doped p(NAGA-*co*-VTZ) hydrogel composite at different scan rates ((5, 10, 20, 50, 100, 200, 300, 400, 500 mV s$^{-1}$) (a), galvanostatic charge-discharge (GCD) curves of the self-healing supercapacitor (b), calculated capacitances at increased current densities (c).

After being thoroughly characterized, the acPPyNTs doped p(NAGA-*co*-VTZ) hydrogels were assembled into a three-electrode cell and electrochemical measurements were performed (Figure 4). **Figure 4a** shows cyclic voltammograms (CVs) of acPPyNTs in 3M KCl at different scan rates (5, 10, 20, 50, 100, 200, 300, 400, 500 mV s$^{-1}$ in the potential window -0.7 to +0.7 V. The CV curves A maximum specific capacitance of 282.62 F g$^{-1}$ was obtained at 5 mV s$^{-1}$. No redox peaks were observed. CGD curves do not show any plateau or battery like-behaviour. The galvanostatic charge-discharge (GCD) curves demonstrated a symmetrical triangular shape (**Figure 4b**). The capacitance diminishes with increasing current, probably because accessibility of ions to the pores of acPPyNTs (**Figure 4c**).

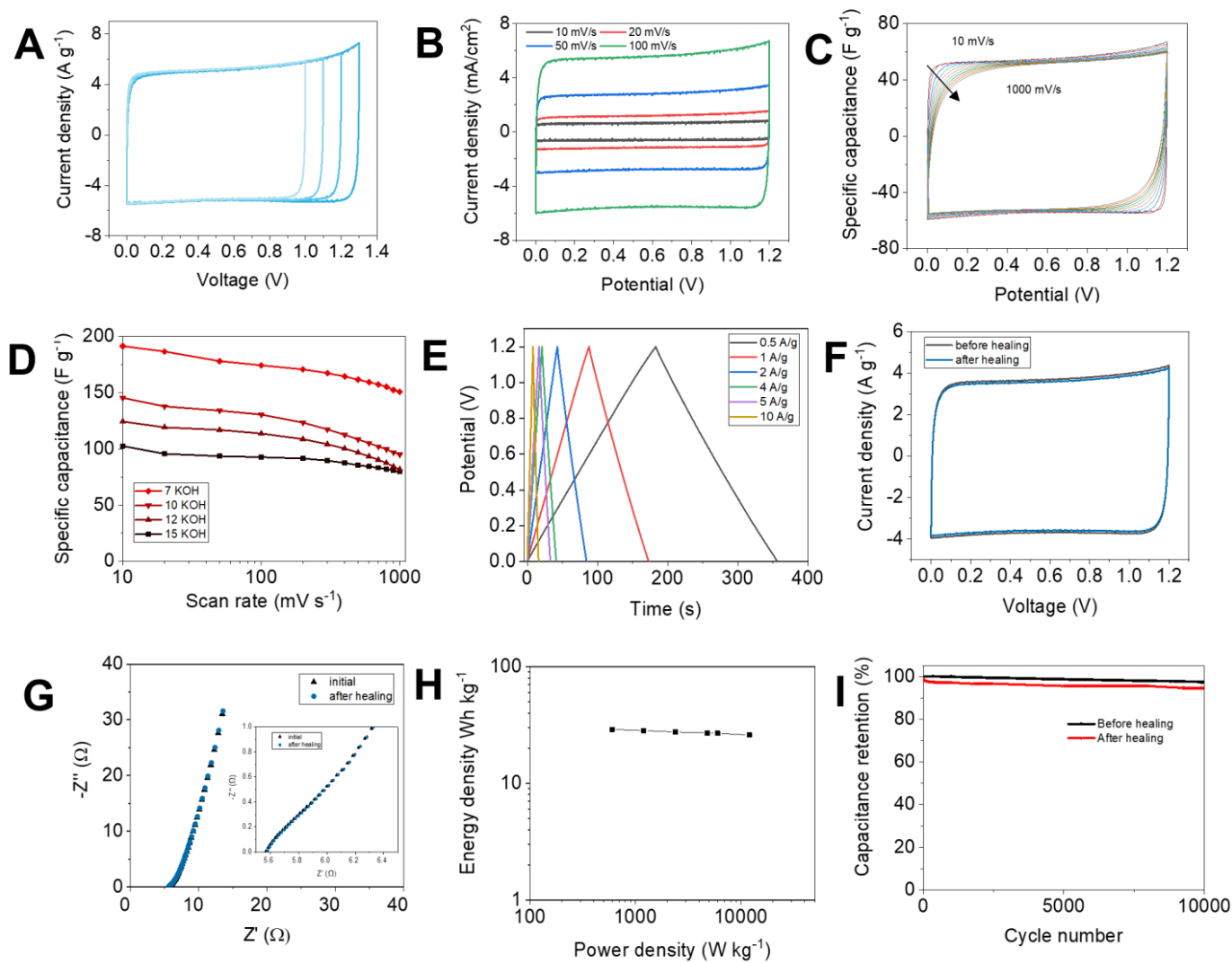

**Figure 5** Cyclic voltammetry (CV) measured in two-electrode system at a scan rate 100 mV s$^{-1}$ at increasing cell voltage (A), CV measurements at lower rates (B) CVs of symmetrical cell in 1.2 V window measured at 5, 10, 20, 50, 100, 200, 300, 400, 500, 600, 700, 800, 900, 1000 mV s$^{-1}$, (C) galvanostatic charge-discharge curves at different current densities and different concentrations of KOH used for activation of carbonized polypyrrole nanotubes (D), galvanostatic charge-discharge (GCD) curves of the self-healing supercapacitor at different current densities (E), CV measurements of p(NAGA-*co*-VTZ) hydrogel-based supercapacitor before and after self-healing (F), impedance spectra of the material (G), Ragon plots of energy density for various samples (H), cycling stability analysis of pristine supercapacitor and after self-healing up to 10 000 cycles (I).

Next, two-electrode measurements were performed. The measurements revealed fast electrochemical response and pure capacitive behaviour of acPPyNTs loaded p(NAGA-*co*-VTZ) supercapacitor. The as prepared polymer hydrogels decorated with acPPyNTs and saturated in 3M KCl were assembled in symmetrical supercapacitor. The use of neutral solvent as strongly required according to the nature of p(NAGA-*co*-VTZ) hydrogel to avoid hydrolysis of the polymer.[27] In our cases, dissolution of p(NAGA-*co*-VTZ) hydrogel film was observed immediately after immersion in e.g. 6 M KOH aqueous solution. To evaluate potential windows, measurements from 0-1, 0-1.1, 0-1.2 and 0-1.3 V at 100 mV s$^{-1}$ were performed. Our measurements determined an excellent capacitive performance of the samples in a working potential window of ~1.2 V. A slight deviation from rectangular shape at 1.1-1.3 V can be attributed to a redox process.

The CV at lower rates is given on **Figure 5B**. From this graph it is evident that no oxidation peaks were observed. The gradual CV curve deviation from the rectangular shaped increase with can rate up to 1000 mV s$^{-1}$ (**Figure 5C**). In addition, the capacitance to 78.4 % at 1000 mV s$^{-1}$. This can be attributed to diffusion-controlled ionic transport at high scan rates. Next, the effect of different KOH concentration, used for cPPyNTs activation was analysed (**Figure 5D**). Based on the capacitance measurement results, the best outcomes were demonstrated when we used 7 M KOH solution with a specific amount of ethanol for activation of the cPPyNTs. After activation with 7 M KOH, in two-electrode composition the highest capacitance of 191.4 F g$^{-1}$ was observed at 10 mV s$^{-1}$. The galvanostatic charge-discharge (GCD) curves demonstrated a symmetrical triangular shape from 0 to 1.2 V in two-electrode assembling **Figure 5E.** GCD demonstrated exception capacitive properties of the material with almost linear change in voltage over time. After self-healing the material demonstrates only slight changes in capacitance (**Figure 5F)**. The Nyquist plots for the acPPyNTs are shown in the **Figure 5G**. The x-intercept in the high frequency region showed equivalent series resistance (ESR) of 5.54 Ω for pristine p(NAGA-*co*-VTZ) film and 5.58 Ω after self-healing, with almost unchanged slope. The ESR value is typical for hydrogel electrolytes. These data showed just small decrease in capacitance after the self-healing. The energy densities versus power densities for as-prepared supercapacitor is presented on **Figure 5H.** For pristine acPPyNTs loaded p(NAGA-*co*-VTZ) hydrogel, the highest energy density of 28.97 Wh kg$^{-1}$ was observed at a power density of 600 W kg$^{-1}$. This performance have superior values compared to many N-doped carbon materials. The power density slightly decreases at higher values of power density: at power density 12 000 W kg$^{-1}$ the value of energy density was 26 Wh kg$^{-1}$. Cyclic stability another essential parameter only 3 % capacitance loss was observed after 10 000 cycles (Figure 5I). Compared to pristine (non-healed supercapacitor a decrease in capacitance was observed after first 500 cycles) (Figure 5I). Nevertheless, after self-healing the supercapacitor capacitance decrease only to 94 % after 10 000 cycles was measured.

The comparison of acPPyNTs-doped p(NAGA-*co*-VTZ) hydrogel with other self-healing materials reported in the literature is given in **Table 1**. As is evident, we obtained an excellent ratio of capacitance and mechanical properties. The proposed acPPyNTs-doped p(NAGA-*co*-VTZ) hydrogel also can be used in wide operation windows of 0-1.2 V. The additional benefit of our material is simplicity of self-healing, which can be performed by heating. Although, some of materials have superior properties in some parameters, they fall short in others. Overall, our material demonstrating high capacitance values, good mechanical properties, a wide potential window, can be considered as an interesting alternative to existing concepts. acPPyNTs-doped p(NAGA-*co*-VTZ) hydrogel does not require addition cross-linking or use of nanoparticles and can be fast healed by heating.

Table 1. Comparison of supercapacitor with self-healing properties

| Material | Capacitance | Potential window | Healing mechanism | Mechanical strength | Strain, % | Publication year/reference |
|---|---|---|---|---|---|---|
| **This work: acPPyNTs loaded p(NAGA-co-VTZ) hydrogel** | **316.86 mF cm$^{-2}$ at 10 mV s$^{-1}$** | **0 to 1.2 V** | **Hydrogen bonding, thermal** | **0.9 MPa** | **1300 %** | **\*** |
| Poly(acrylic acid), crosslinked by N, N'-bis(acryloyl)cystamine-coated Au nanoparticles (AuNP@BACAs). | 20.4 mF cm$^{-2}$ | 0 to 2 V | Thiol-Au nanoparticles | 17.6 kPa | 750% | 2023[15] |
| BACA/Fe$_3$O$_4$@Au/polyacrylamide | 1264 mF cm$^{-2}$ | 0 to 0.8 V | Thiol- Fe$_3$O$_4$@Au NPs | 3.1 MPa | 2250 % | 2021[2] |
| Copolymer of vinylimidazole and hydroxypropyl acrylate | 93.2 F g$^{-1}$ at 45 °C | 0 to 1 V | Hydrogen bonding | ~8 kPa | ~140 % | 2023[28] |
| Poly(2-acrylamido-2-methyl-1-propanesulfonic acid)/polyvinylimidazole | 106 mF cm$^2$ at 50 mV/s | 0 to 0.8 V | Electrostatic interaction and hydrogen bonding | 154.57 kPa, | 1137% | 2021[29] |
| Gelatin methacrylate cellulose nanocrystal, tannic acid polyaniline and reduced graphene oxide | 1861.21 mF cm$^{-3}$ | 0 to 0.8 V | Treatment with tannic acid for 24 h | 17.42 kPa | 71.11% | 2020[30] |
| Carboxylated polyurethane (PU) MXene−rGO aerogel | 34.6 mF cm$^{-2}$ at a scan rate of 1 mV s$^{-1}$ | 0 to 0.6 V | Hydrogen bonding | 10.24 kPa | 70% | 2018[31] |
| Physically cross-linked poly(vinyl alcohol)−H$_2$SO$_4$+ PANI | 25.86 mF cm$^{-2}$ at a current density of 0.05 mA cm$^{-2}$ | −0.2 to 0.8 V | Hydrogen bondings 94% capacitance 2000 cycles | 440 kPa | ~350% | 2020[32] |

.

## 3. Conclusion

In summary, we prepared high strength self-healable supercapacitor based on acPPyNTs doped p(NAGA-*co*-VTZ) supramolecular polymer with upper critical solution temperature. This material does not require additional crosslinking and demonstrated self-healing ability through hydrogen bonding at 95 °C during 1 min. This supercapacitor demonstrates high specific 282.62 F g$^{-1}$ at 0.2 A g$^{-1}$ and areal capacitance 316.86 mF cm$^{-2}$ at scan rate of 10 mV s$^{-1}$, in 0-1.2 V voltage window in aqueous 3 M KCl electrolyte, shows high energy density and cyclic performance (capacitance retention of 97 % after 10 000 cycles and 94 % after self-healing and 10 000 CVA cycles). SEM analysis in combination with mechanical tests showed that hydrogel mechanical properties were strongly dependent on initial monomer concentration. The prepared composite demonstrate outstanding mechanical properties and demonstrated resistance to cut. The prepared material have a high potential in the field of flexible electronics.

## 4. Experimental Section

*Chemicals:* Glycinamide hydrochloride (98 %, abcr), acryloyl chloride (96 %, abcr), hydroxy-2-methyl-1-phenyl-1-propanone (97 %, Merck), and 1-vinyl-1,2,4-triazole (VTZ ≥ 97%, Merck), pyrrole, iron (III) chloride, diethyl ether, potassium carbonate, acetone, ethanol 99.5 % and potassium chloride were obtained from PENTA.

*Synthesis of polypyrrole nanotubes (PPyNTs):* PPyNTs were synthesized using modified procedure.[11] Typically, 2.43 g of FeCl$_3$ was dissolved in 300 ml of 5 mM solution of methyl orange (MO). The mixture was stirred on ice bath and then, 1.05 ml of pyrrole was added. After 15 min the solution became black, indicating the polymerization of pyrrole, the solution was left mixing for 24 h to complete polymerization. The precipitate was washed with deionized water and ethanol several times, then placed into Soxhlet extractor (acetone) for 2 weeks to remove the excess of MO. After 2 weeks the PPyNTs were dried and used carbonized and activated.

*Carbonization and activation of PPyNTs:* Next, polypyrrole nanotubes were carbonized under nitrogen flow using 5 °C/min heating to 1000 °C under nitrogen flow. The carbonized PPyNTs were then placed into plastic dishes and different 7 M, 10 M, 12 M, 15 M KOH water solutions were used. To achieve wettability of the carbonized PPyNTs nanotubes 2 ml of ethanol were added to the solutions. The solutions were rigorously stirred for 20 h, filtered and dried at 80 °C overnight. After this the samples were placed into tube furnace and carbonized in vacuum at 750 °C using 5°C/min heating.

*Synthesis of N-acryloyl glycinamide monomer:* N-acryloyl glycinamide has been prepared according to the modified route.[26] In a 250 ml three-neck round bottom flask equipped with mechanical stirrer glycinamide hydrochloride (5.78 g, 52.25 mmol) were dissolved in 32 ml of deionized water. The solution was cooled in ice bath and acryloyl chloride (4.2 ml, 51.25 mmol) dissolved in 62 ml diethylether was added dropwise over 30 min with fast stirring 300 rpm. The suspension was further stirred at RT for 2 h. Diethylether was removed by rotary evaporation at 35 °C. The remaining aqueous phase was freeze dried. The crude brittle solid was extracted with acetone (6 times, 125 mL, 40 °C, stirring for at least 15 min. Potassium salts were filtered off and the acetone was removed by rotary evaporation at 35 °C). The $^1$H NMR and $^{13}$C NMR results are presented below. $^1$H NMR (500 MHz, D$_2$O) δ 6.26 (dd, $J$ = 17.1, 10.1 Hz, 1H), 6.18 (dd, $J$ = 17.1, 1.4 Hz, 1H), 5.75 (dd, $J$ = 10.1, 1.4 Hz, 1H), 4.72 (s, 2H), 3.92 (s, 2H). $^{13}$C NMR (126 MHz, D$_2$O) δ 174.27, 169.06, 129.37, 128.20, 42.09.

*Synthesis of p(NAGA-co-VTZ) polymer:* Polymer was synthesized by photopolymerization under borosilicate glass using spacers. According to our experiments, the optimal spacer size was 500 μm. First, different amount of NAGA monomer and VTZ were dissolved into deionized water to obtain 9, 16, 23, 30, 37 wt % solution concentration. Then gel was soaked in 3M KCl.

*Preparation of supercapacitor and electrochemical measurements:* For two-electrode configuration, cell supercapacitors were assembled by sandwiching of acPPyNTs and p(NAGA-*co*-VTZ) as separator. Next the sample was soaked in 3M KCl. 3M KCl was also used as the electrolyte for supercapacitor. The specific capacitance was calculated according to formula:

$$C_s = 2(I/m)(\Delta t/\Delta V) \tag{1}$$

where I/m is current density (A g$^{-1}$), Δt is discharge time (s), ΔV is potential window, C is capacitance of supercapacitor. The energy density (E) and power density (P) were calculated according to equations:

$$E = C(\Delta V)^2/2 \times 3.6 \tag{2}$$

$$P = E/\Delta t \tag{3}$$

**Acknowledgement**

This work was supported by the Czech Science Foundation (GA CR) under the project no. 22-25734S